\documentclass[journal,twoside,web]{ieeecolor}
\usepackage{generic}
\usepackage{cite}
\usepackage{amsmath,amssymb,amsfonts}
\usepackage{algorithmic}
\usepackage{graphicx}
\usepackage{textcomp}
\usepackage{color}
\usepackage{soul}
\usepackage{textcomp}
\def\BibTeX{{\rm B\kern-.05em{\sc i\kern-.025em b}\kern-.08em
 T\kern-.1667em\lower.7ex\hbox{E}\kern-.125emX}}
\begin{document}
\title{Developing RPC-Net: Leveraging High-Density Electromyography and Machine Learning for Improved Hand Position Estimation} \author{Giovanni Rolandino, \IEEEmembership{Student Member, IEEE}, Marco Gagliardi, Taian Martins,\\ Giacinto Luigi Cerone, \IEEEmembership{Member, IEEE} , Brian Andrews and James J. FitzGerald, \IEEEmembership{Member, IEEE}  \thanks{G. Rolandino, J. J. FitzGerald and B. Andrews are with the Nuffield Department of Surgical Sciences,
University of Oxford, Oxford, UK (e-mail: giovanni.rolandino@nds.ox.ac.uk). M. Gagliardi, T. Martins and G. L. Cerone are with LISiN (Department of Electronics and Telecommunications), Politecnico di Torino, Turin, Italy (e-mail: taian.martins@polito.it). This article has supplementary downloadable material available at https://zenodo.org/records/10000899, provided by the authors. Copyright (c) 2023 IEEE. Personal use of this material is permitted. However, permission to use this material for any other purposes must be obtained from the IEEE by sending an email to pubs-permissions@ieee.org.}}

\maketitle

\begin{abstract}
\textit{Objective:} The purpose of this study was to develop and evaluate the performance of RPC-Net (Recursive Prosthetic Control Network), a novel method using simple neural network architectures to translate electromyographic activity into hand position with high accuracy and computational efficiency. \textit{Methods:} RPC-Net uses a regression-based approach to convert forearm electromyographic signals into hand kinematics. We tested the adaptability of the algorithm to different conditions and compared its performance with that of solutions from the academic literature. \textit{Results:} RPC-Net demonstrated a high degree of accuracy in predicting hand position from electromyographic activity, outperforming other solutions with the same computational cost. Including previous position data consistently improved results across subjects and conditions. RPC-Net showed robustness against a reduction in the number of electromyography electrodes used and shorter input signals, indicating potential for further reduction in computational cost. \textit{Conclusion:} The results demonstrate that RPC-Net is capable of accurately translating forearm electromyographic activity into hand position, offering a practical and adaptable tool that may be accessible in clinical settings. \textit{Significance:} The development of RPC-Net represents a significant advancement. In clinical settings, its application could enable prosthetic devices to be controlled in a way that feels more natural, improving the quality of life for individuals with limb loss.
\end{abstract}

\begin{IEEEkeywords}
Artificial Neural Networks, electromyography, Machine Learning, Prosthetic hand.
\end{IEEEkeywords}

\section{Introduction}
\label{sec:Intro}

\IEEEPARstart{I}{n} 2019, the prevalence of upper limb amputations in the United States, United Kingdom, and the European Union was more than 1,000,000 (incidence 40,000), 100,000 (incidence 3,200), and 800,000 (incidence 25,000), respectively \cite{WinNT}. Loss of upper limb functionality has a major impact on quality of life, underscoring the pressing need for effective solutions \cite{mohammed2014quality}. Great strides have been made in designing advanced upper limb prostheses and developing control strategies for these devices. Specifically, efforts have been directed towards creating prostheses that replicate the kinematic complexity of the human hand \cite{carrozza2006design}\cite{catalano2014adaptive}\cite{controzzi2016sssa}\cite{laffranchi2020hannes}\cite{fajardo2020galileo}\cite{kyberd2017assessment}. Many types of control solutions for these devices have been explored, with a primary focus on surface electromyography (EMG), currently the most viable option for prosthetic control \cite{jiang2012myoelectric}\cite{d2023online}.

Despite the considerable advancements in technology, abandonment rates for prosthetic devices, a good indicator of user satisfaction, are consistently high, and are rarely below 30\% \cite{biddiss2007up}\cite{smail2021comfort}\cite{salminger2022current}. Factors contributing to abandonment include comfort, function, and appearance \cite{smail2021comfort}, and it appears that none of those are strongly predominant over the others \cite{kyberd2011survey}. In general, users feel the prosthesis does not address their needs \cite{ostlie2012prosthesis}, often find them uncomfortable or painful and feel that control is not natural enough \cite{cordella2016literature} \cite{biddiss2007upper}. The definition of control naturalness varies in the literature, encompassing factors such as independent finger movement, force control capability, ease in performing daily tasks, and the inclusion of sensory feedback \cite{biddiss2007up} \cite{smail2021comfort}. Overall, the fact that technical developments have not led to visible improvements in abandonment rates or consumer satisfaction suggests that current approaches do not meet patient needs \cite{salminger2022current}.

An analysis of current control solutions may provide insight into the factors contributing to this trend. Machine Learning (ML) and its subset Deep Learning (DL) have been extensively employed in recent years to address prosthetic control. Applications of classification-based algorithms, a subset of ML, have been investigated \cite{mendez2021}\cite{Ciancio2016}\cite{Iqbal2018}\cite{Cloutier2013}\cite{ahmadizadeh2021human}. This type of algorithm predicts discrete or categorical output labels and maps input features to one of a finite number of classes or categories. For prosthetic control, this entails classifying electromyographic signals as one of several predefined movements, each of which is assumed to be consistent. Research has been focused on developing more complex solutions to improve accuracy in the classification of more and more classes \cite{atzori2016deep}\cite{hu2018novel}. Classification-based approaches, however, deviate significantly from how humans control their hands, and thus, may not be able to provide users with the natural-feeling control they desire \cite{biddiss2007consumer}. 

In contrast to classification approaches, regression-based solutions offer a more natural control mechanism, taking steps towards human-like prosthetic control \cite{vujaklija2023prosthetics}\cite{simpetru2022accurate}\cite{nowak2022simultaneous}\cite{thomas2022utility}\cite{george2019biomimetic}. This type of solution has been, in recent years, object of growing interest because of its potential \cite{Castellini2008}\cite{mendez2021}\cite{cimolato2020hybrid} \cite{choo2023use}. Despite the clear improvement that regression-based approaches bring, some problems persist. First, most existing solutions do not control the complete kinematics of the hand, instead focusing solely on either the wrist or the fingers, even if control of many kinematic Degrees of Freedom (DoFs) independently was identified as an important attribute \cite{jiang2012myoelectric}\cite{smail2021comfort}. Second, some of these solutions require electromyogram signals acquired distally. However, this is incompatible with the needs of upper limb amputees; only 5\% of hand amputees retain distal muscles in their forearms, whereas more than 30\% of the same population retain proximal forearm muscles \cite{cordella2016literature}. Thirdly, most of these solutions employ computationally intensive algorithms with substantial computational cost, which may render them less suitable for embedding into devices or application in clinical settings, ultimately hindering their practical applicability \cite{atzori2015control}\cite{castellini2009multi}\cite{de1997use}\cite{tenore2008decoding}\cite{fukuda2003human}.

\begin{figure}[t]
 \centerline{\includegraphics[width=\columnwidth]{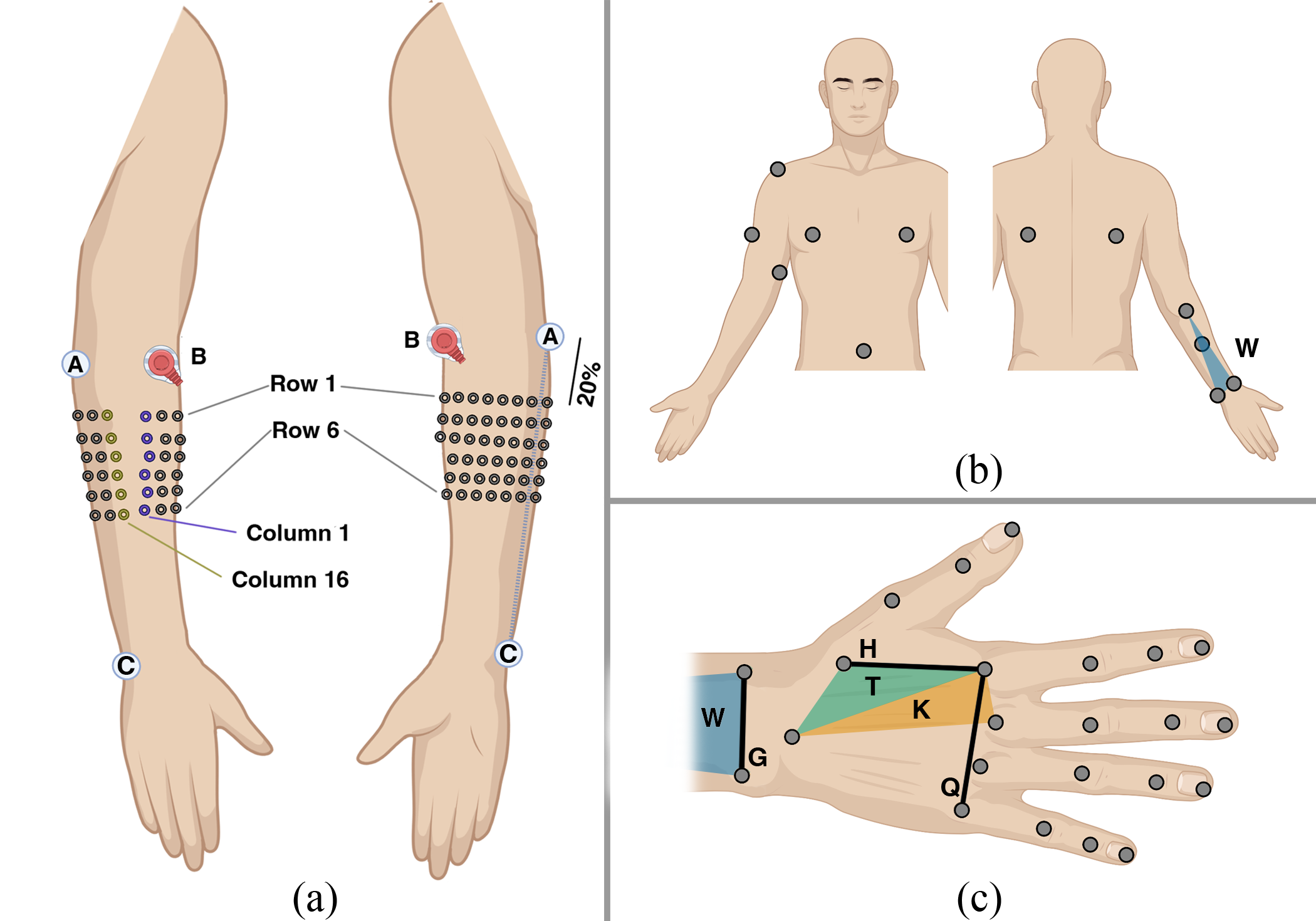}}
 \caption{(a) Position of electrodes on the forearm of the subject. Ventral (L) and dorsal (R) view (A: Medial epicondyle, B: Lateral epicondyle, C: Pisiform bone). (b) Placement of 12 infrared markers on the body of the subject. Plane W is highlighted in blue. (c) Placement of 21 infrared markers on the hand (and 2 forearm markers, also depicted in (b)). Markers are placed proximally to each joint. Planes K and T highlighted in orange and green respectively. Q, H and G lines in black.}
 \label{fig:big_fig}
\end{figure}

We suggest that these limitations are contributing to the observed low satisfaction among prosthesis users. To address the shortcomings observed, and confident that accurate hand position estimation is crucial for improving prosthetic device control, we developed RPC-Net (Recursive Prosthetic Control Network). This novel, computationally efficient, deep network leverages regression principles to estimate hand position (as a high-DoF kinematic model) from electromyographic activity, and is intended to be implemented as control solution for articulated hand prostheses. Regression-based control and a high-DoF kinematic model address naturalness of control, a key factor in device abandonment \cite{vujaklija2023prosthetics}\cite{simpetru2022accurate}\cite{nowak2022simultaneous}\cite{thomas2022utility}\cite{george2019biomimetic}. RPC-Net is recursive (making use of previous estimates to refine the current one) and uses 96 high-density surface EMG (HD-sEMG) channels from the proximal forearm as source signal, fewer than most control approaches and in accordance with the anatomical needs of amputees \cite{Ciancio2016}\cite{mendez2021}.

We hypothesize that RPC-Net, designed to meet the needs of prosthesis users while being computationally efficient, can provide high-quality hand position estimates from the electromyogram and that its performance is superior to state-of-the-art solutions. We further hypothesize that RPC-Net is robust against changes in input signal length and the number of electromyographic channels used, and that information about previous position can improve its performance. In this study, we introduce RPC-Net and validate it to demonstrate these experimental hypotheses. RPC-Net is tested offline and on healthy subjects, focusing on the accuracy of the estimate of the position from the EMG signal. For this purpose, we collected electromyographic and hand position data from twelve healthy subjects, and used RPC-Net to translate the electromyographic data into hand position. This first evaluation is a stepping stone towards the development of a prosthetic solution implementing RPC-Net that satisfies user needs, is efficient, and conducive to a more natural control experience.

\section{Materials and Methods}

\subsection{Subjects and Experimental Protocol}

\subsubsection{Subjects}
Our study included twelve healthy subjects, seven males and five females (aged 20-26, weighing 55-90 kg and 165-195 cm tall). Each participant provided written, informed consent before participation. All subjects were right-handed, had no surgical interventions on their dominant arm, and had a forearm circumference between 20 and 30 cm. The experimental procedures adhered to the Declaration of Helsinki and were approved by the local ethics committee (CER-Polito, Prot. No. 107460/2023).

\subsubsection{Experimental Protocol} \label{sec:task}
The acquisition protocol consisted of two phases. In Phase 1 (P1), a single participant (S0) was involved, whereas Phase 2 (P2) incorporated the remaining eleven participants (S1-S11). The acquisition procedures for both phases were identical, with the only difference being their duration. In P1, twelve trials were included: eleven for training and one for testing. In P2, six trials per subject were considered: five for training and one for testing. The trial to be used for testing was selected randomly. During the trials, high-density EMG and hand position data of the participants were acquired as they transitioned between a set of 18 hand poses. This set comprised 6 wrist poses (flexion; extension; adduction; abduction; pronation; supination), 8 finger poses (index finger metacarpo-phalangeal flexion; index finger metacarpo-phalangeal extension; index finger proximal interphalangeal flexion; index finger proximal interphalangeal extension; flexion of the middle, ring, and little fingers; extension of the middle, ring, and little fingers; adduction of the index and middle fingers; abduction of the index and middle fingers), and 4 thumb poses (flexion; extension; adduction; abduction). Each trial included 54 poses (18 poses repeated 3 times each). Participants were seated with their dominant forearm positioned on a vertical support at shoulder height. Initially, participants were instructed to relax their hands and wrists. Subsequently, a monitor presented one of the 18 hand poses every 8 seconds, in random order to to prevent participants from anticipating the sequence. Participants were given 2 seconds to recognize the pose before transitioning their hand from the previous to the prompted pose and were not instructed to transition at a specific speed. At the end of the trial, the subject was instructed to relax their hands and wrists. Each trial was 450 seconds long.

\subsection{Data Acquisition}

All data have been made available for online access \cite{zenodo_data}.

\subsubsection{High-Density surface EMG data}

EMG was recorded on the surface of the dominant forearm using the MEACS system, the EMG amplifier developed at LISiN (Politecnico di Torino, Turin, Italy) \cite{Cerone2019}\cite{Cerone2021}. The system is made up of multiple Sensor Units (SU), each measuring 34 mm $\times$ 30 mm $\times$ 15 mm and sampling 32 channels at $f_s$=2.048 kHz (192 V/V gain, 16 bit resolution, 2.4 V dynamic range). Three SUs were used, each connected to an anisotropic electrode array (2 rows and 16 columns, with 10 mm and 15 mm inter-electrode distance respectively) for a total of $N$=96 acquired monopolar electromyographic channels. The electrodes were arranged in 6 rows and 16 columns around the circumference of the forearm, covering approximately a third of its length (Fig. \ref{fig:big_fig}). The proximal row of electrodes (row 1) was positioned at 20 \% of the distance between the medial epicondyle and the pisiform bone. The reference electrode was positioned on the lateral epicondyle. The electromyographic signal was used as input for RPC-Net during the phases of training and testing.

\begin{figure}[t]
 \centerline{\includegraphics[width=\columnwidth]{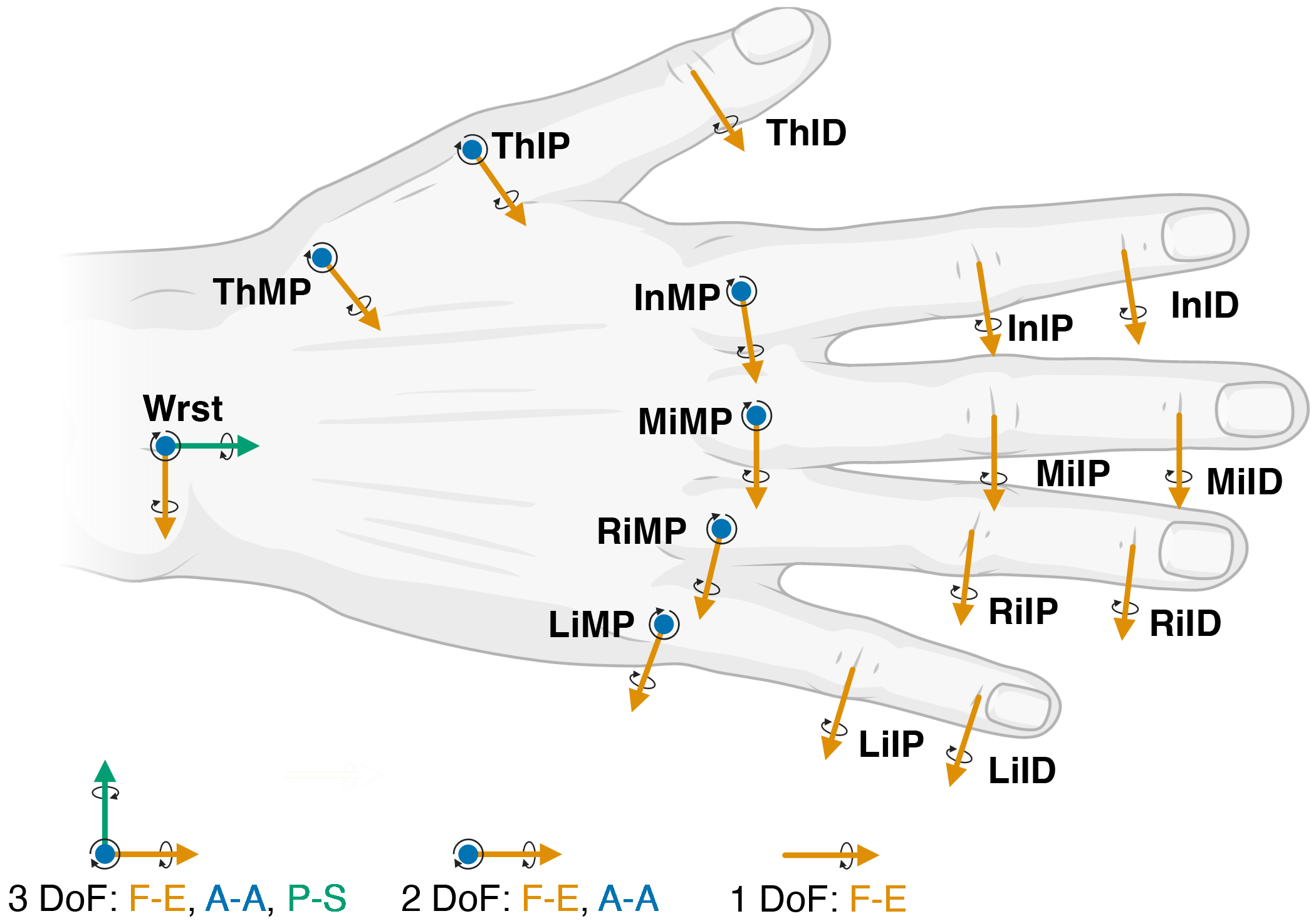}}
 \caption{Kinematic model used for the Inverse Kinematic Algorithm. Each joint of the hand has 1 to 3 kinematic DoFs. Each DoF can be characterized as Flexion-Extension (F-E), Adduction-Abduction (A-A) or Pronation-Supination (P-S), as shown in legend (bottom left). Each joint is given a four letter code. The arrow represents the positioning of the axis of rotation. The circle at the origin represents a rotation axis orthogonal to the back of the hand (dorsal direction). }
 \label{fig:kin_mod}
\end{figure}

\begin{figure*}[t]
 \centering
  \includegraphics[width=\textwidth,height=0.7\textheight,keepaspectratio]{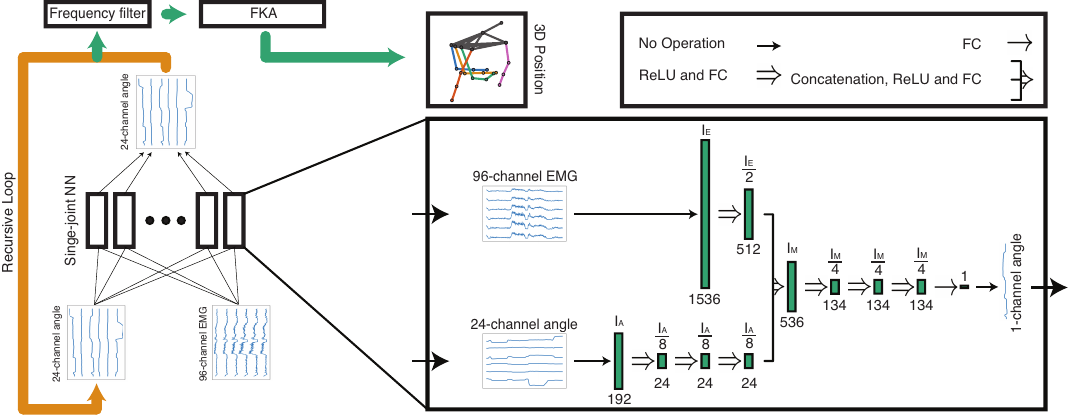}
 \caption{RPC-Net. The whole network made up by 24 individual sub-networks, each of them trained to regress one of the 24 hand joints. The arrow on the left highlights how the output of RPC-Net is used as input for the subsequent iteration. The detail of one of the 24 sub-networks is shown on the right-hand side of the figure. The two branches are visible. "E" refers to the EMG branch and "A" refers to the angle branch. I{\tiny E} and I{\tiny A} represent the width of the EMG (1536 units) and angle (192 units) inputs respectively. I{\tiny M}=I{\tiny E}+I{\tiny A} is the width of the merged outputs of the two branches (536 units). The value on top of the green layers indicate the width of the corresponding layer as a function of I{\tiny x}, the input size of that branch, expressed numerically beneath each layer. The top left part of the figure displays the filtering and conversion to 3D part that happen outside the network loop.}
 \label{fig:NN_2}
\end{figure*}

\begin{figure}[t]
 \centerline{\includegraphics[width=\columnwidth]{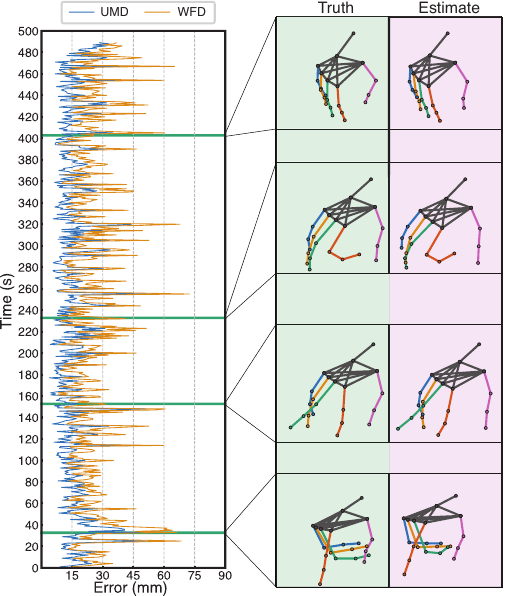}}
 \caption{Estimation error. The two lines show the UMD and the WFD for the test trial of S0. For selected time points, the estimated position (right) and the actual one (left) are shown on the side.}
 \label{fig:big_pic}

\end{figure}

\begin{figure*}[t]
 \centering
  \includegraphics[width=\textwidth,height=0.7\textheight,keepaspectratio]{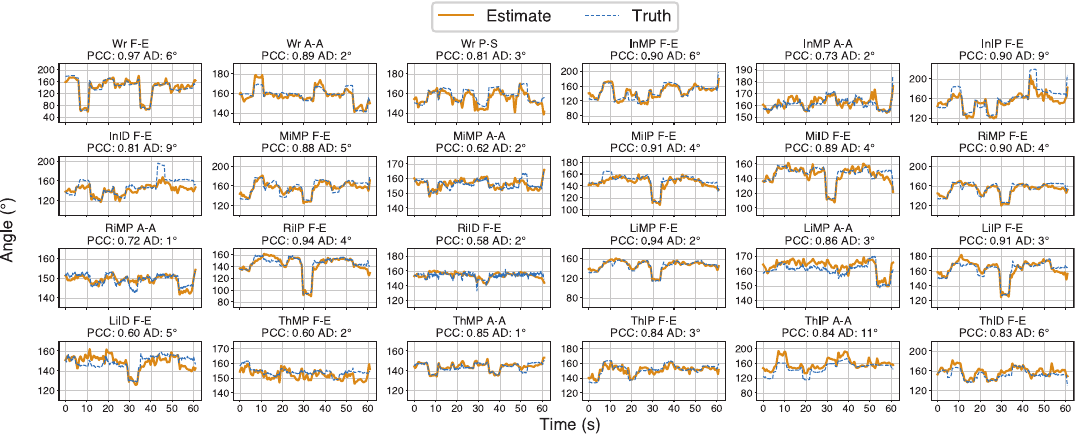}
 \caption{Comparison between estimated joint angle and actual angle. The 24 joints considered are shown. At the top of each subplot the abbreviation of the corresponding joint and DoF, as defined in Fig. \ref{fig:kin_mod}. Two measures of accuracy are also included: PCC between actual and estimated position and the Mean Angular Distance (MAD), in degrees, between the values over time.}
 \label{fig:results_small}

\end{figure*}

\begin{figure*}[t]
 \centering
  \includegraphics[width=\textwidth,height=0.7\textheight,keepaspectratio]{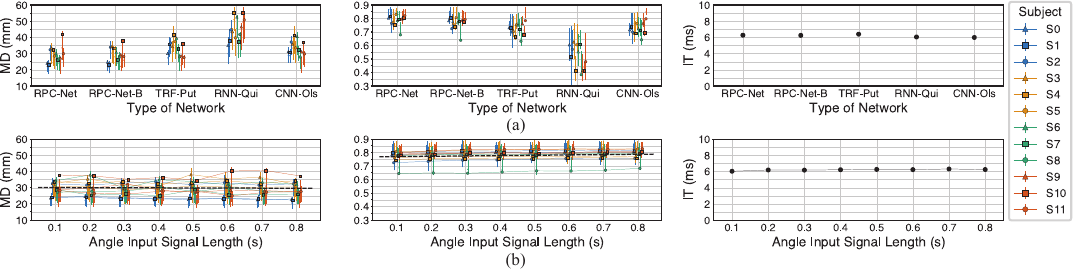}

 \caption{(a) Comparison of the performance between RPC-Net, RPC-Net-B, TRF-Put, RNN-Qui, and CNN-Ols. The figure reflects the experimental conditions outlined in section \ref{sec:thrsltns} and in section \ref{sec:pr_j}. one-sided paired t-test results for MD (H0: RPC-Net $\geq$ RPC-Net-B): $t(11)$=-0.96 $p$=1.7e-01, (H0: RPC-Net $\geq$ TRF-Put): $t(11)$=-2.78 $p$=8.9e-03, (H0: RPC-Net $\geq$ RNN-Qui): $t(11)$=-11.2 $p$=1.2e-07, (H0: RPC-Net $\geq$ CNN-Ols): $t(11)$=-3.55 $p$=2.3e-03. one-sided paired t-test results for MPCC (H0: RPC-Net $\leq$ RPC-Net-B): $t(11)$-6.88 $p$=1.3e-05, (H0: RPC-Net $\leq$ TRF-Put): $t(11)$=-8.75 $p$=1.4e-06, (H0: RPC-Net $\leq$ RNN-Qui): $t(11)$=-11.3 $p$=1.1e-07, (H0: RPC-Net $\leq$ CNN-Ols): $t(11)$=-9.19 $p$=8.5e-07. (b) Performance as a function of angle branch input signal length, as described in section \ref{sec:l_o_i}. Regression analysis results for MD (H0: $\beta_{1}$=0): $\hat{\beta_{1}}$=-0.057 mm/s, $SE$=2.0e-01, $t$=-2.9e-01, $p$=7.8e-01, $R^{2}$=1.0e-03, adjusted $R^{2}$=-1.0e-02. Regression analysis results for MPCC (H0: $\beta_{1}$=0): $\hat{\beta_{1}}$=0.002 1/s, $SE$=2.0e-03, $t$=1.3e-00, $p$=2.0e-01, $R^{2}$=1.7e-02, adjusted $R^{2}$=-7.0e-03.} \label{fig:results_previous_state}

\end{figure*}

\begin{figure*}[t]
 \centering
  \includegraphics[width=\textwidth,height=0.7\textheight,keepaspectratio]{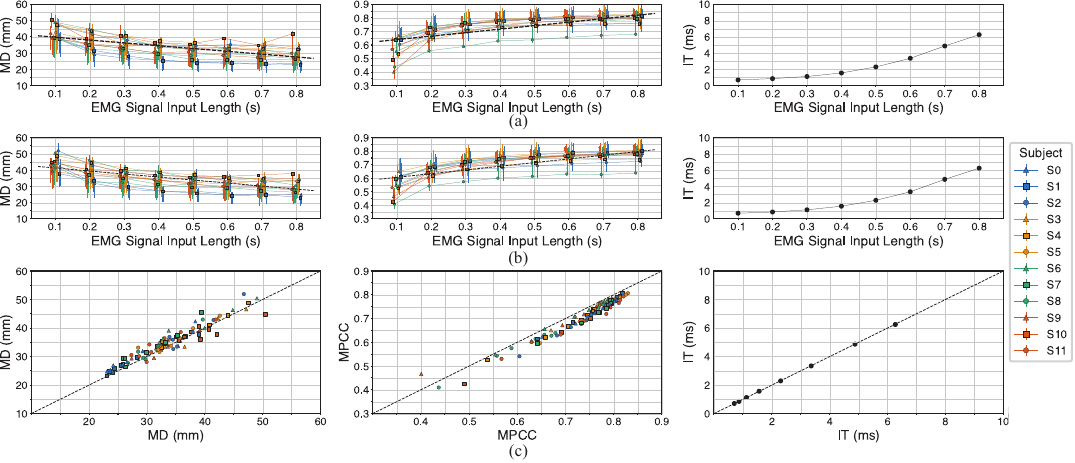}
 \caption{Performance as a function of EMG branch input signal length for RPC-Net and RPC-Net-B, as defined in section \ref{sec:l_o_i}. (a) RPC-Net. Regression analysis results for MD (H0: $\beta_{1}$=0): $\hat{\beta_{1}}$=-17.362 mm/s, $SE$=2.2e-00, $t$=-8.0e-00, $p$=2.4e-12, $R^{2}$=4.0e-01, adjusted $R^{2}$=4.0e-01. Regression analysis results for MPCC (H0: $\beta_{1}$=0): $\hat{\beta_{1}}$=0.257 1/s, $SE$=2.8e-02, $t$=9.3e-00, $p$=4.6e-15, $R^{2}$=4.8e-1, adjusted $R^{2}$=4.8e-1. (b) RPC-Net-B. Regression analysis results for MD (H0: $\beta_{1}$=0): $\hat{\beta_{1}}$=-18.535 mm/s, $SE$=2.0e+00, $t$=9.4e-00, $p$=3.9e-15, $R^{2}$=4.8e-1, adjusted $R^{2}$=4.8e-1. Regression analysis results for MPCC (H0: $\beta_{1}$=0): $\hat{\beta_{1}}$=0.272 1/s, $SE$=2.6e-02, $t$=1.0e+01, $p$=2.5e-17, $R^{2}$=5.4e-1, adjusted $R^{2}$=5.3e-1. (c) Comparative analysis between RPC-Net and RPC-Net-B. Wilcoxon signed-rank test results for MD (H0:median RPC-Net $\geq$ median RPC-Net-B): $W$=786 $p$=8.8e-09. Wilcoxon signed-rank test results for MPCC (H0:median RPC-Net $\leq$ median RPC-Net-B): $W$=96, $p$=1.7e-16. }
\label{fig:results_depth_all}
 
\end{figure*}

\begin{figure*}[t]
 \centering
  \includegraphics[width=\textwidth,height=0.7\textheight,keepaspectratio]{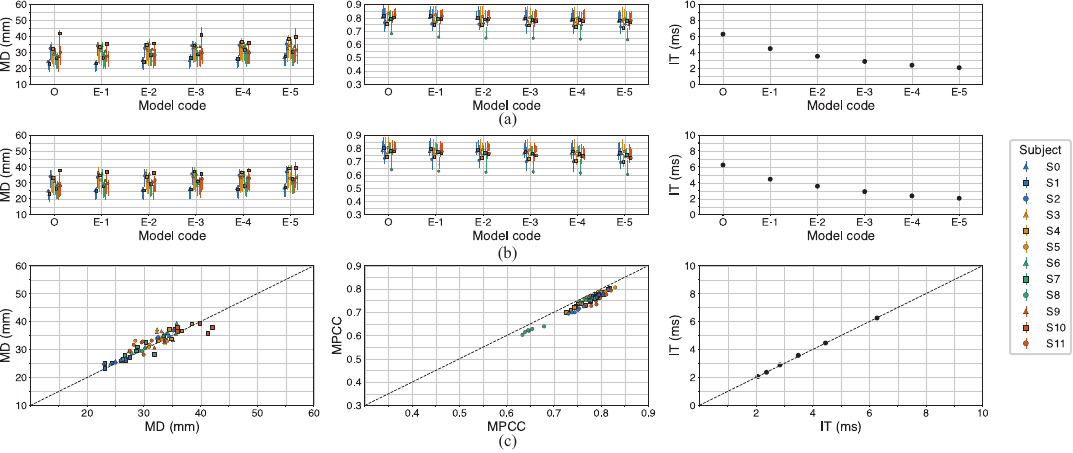}
 \caption{Performance as a function of EMG branch layer width for RPC-Net and RPC-Net-B. The x-axis shows the code defined in \ref{sec:d_o_i}. (a) RPC-Net. (b) RPC-Net-B. (c) Comparative analysis between RPC-Net and RPC-Net-B. Wilcoxon signed-rank test results for MD (H0:median RPC-Net $\geq$ median RPC-Net-B): $W$=577 $p$=1.8e-05. Wilcoxon signed-rank test results for MPCC (H0:median RPC-Net $\leq$ median RPC-Net-B): $W$=0, $p$=8.3e-14.}
 \label{fig:results_fcndepth_all}
\end{figure*}

\begin{figure*}[t]
 \centering
  \includegraphics[width=\textwidth,height=0.7\textheight,keepaspectratio]{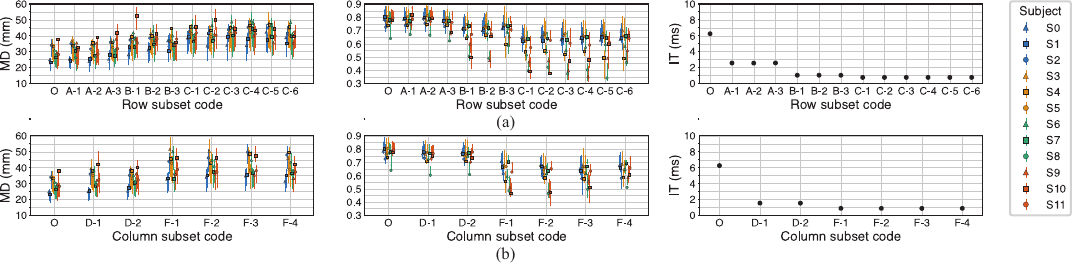}
\caption{Performance as a function of the subset of electrodes used, as described in section \ref{sec:n_o_e}. (a) Rows subsets. (b) Columns subsets. }\label{fig:nr_electrodes}
\end{figure*}

\begin{figure*}[t]
 \centering
  \includegraphics[width=\textwidth,height=0.7\textheight,keepaspectratio]{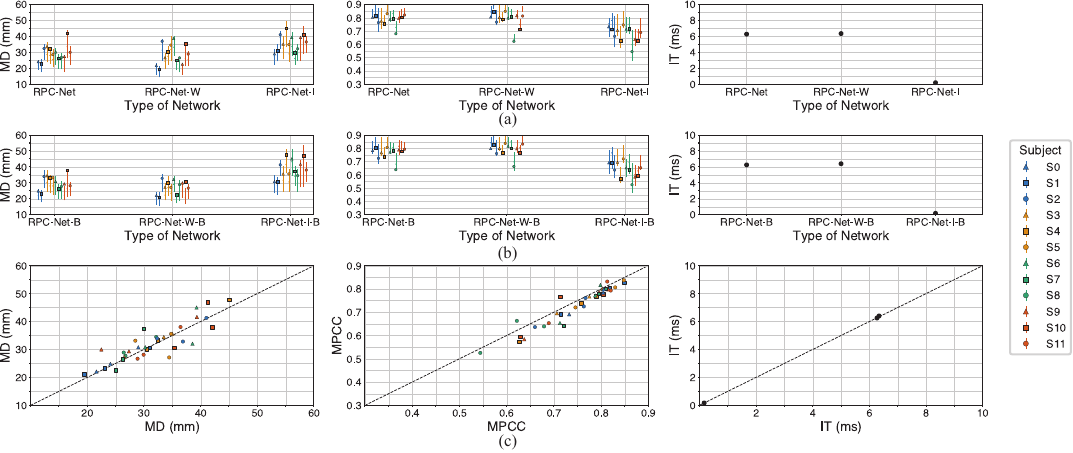}
\caption{Comparison between RPC-Net-W, RPC-Net-I, and RPC-Net, as described in section \ref{sec:s_o_24}. (a) RPC-Net. one-sided paired t-test results for MD (H0: RPC-Net $\geq$ RPC-Net-I): $t(11)$=0.609 $p$=7.2e-01, (H0: RPC-Net $\geq$ RPC-Net-W): $t(11)$=-5.709 $p$=1.6e-05. one-sided paired t-test results for MPCC (H0: RPC-Net $\leq$ RPC-Net-I): $t(11)$=0.025 $p$=5.1e-01, (H0: RPC-Net $\leq$ RPC-Net-W): $t(11)$=-10.412 $p$=2.5e-07. (b) RPC-Net-B. one-sided paired t-test results for MD (H0: RPC-Net-B $\geq$ RPC-Net-I-B): $t(11)$=3.069 $p$=9.9e-01, (H0: RPC-Net-B $\geq$ RPC-Net-W-B): $t(11)$=-7.019 $p$=1.1e-05. one-sided paired t-test results for MPCC (H0: RPC-Net-B $\leq$ RPC-Net-I-B): $t(11)$=5.471 $p$=9.8e-01, (H0: RPC-Net-B $\leq$ RPC-Net-W-B): $t(11)$=-10.682 $p$=1.9e-07. (c) Comparative analysis between RPC-Net and RPC-Net-B. Sign test results for MD (H0:median RPC-Net $\geq$ median RPC-Net-B): $W$=225, $p$=4.5e-02. Sign test results for MPCC (H0:median RPC-Net $\leq$ median RPC-Net-B): $W$=92, $p$=7.6e-05.}\label{fig:separate_or_not}
\end{figure*}

\subsubsection{Hand position data}

Hand position data were acquired using a motion capture system (VICON Motus; VICON Motion Systems, Centennial, Oxford, UK) sampling at 100 samples/s. The setup included 12 infrared cameras (Vero v2.2). A total of $M_{h}$=21 infrared reflective markers (diameter of 6 mm) were positioned on the dominant hand of the subject, embedded in a glove. Additionally, 12 markers were placed on the upper limb and trunk, resulting in $M$=33 markers in total (Fig. \ref{fig:big_fig}). The hand position data, translated to joint angles using the Inverse Kinematic Algorithm (IKA) defined below, was used both as input and as target value for the training phase of RPC-Net.

\subsection{Inverse Kinematic Algorithm}

We developed an Inverse Kinematic Algorithm (IKA) to translate 3D marker positions, as captured by the motion capture system, into hand joint angles, the selection of which is informed by a 24-DoF kinematic model (Fig. \ref{fig:kin_mod}) proposed by Lee et al. \cite{lee1995model}. Although the original model defines 27 kinematic DoFs, we excluded the three associated with spatial origin positioning, as the position of the wrist was fixed in our protocol. The core of the IKA is the optimization process designed to identify the 24 joint angles (one per kinematic DoF) that best approximate the marker positions. The process is executed for each frame captured by the VICON system. The algorithm comprises three phases: 1) wrist angles estimation (3 angles), 2) finger angles estimation (4 angles per finger) and 3) thumb angles estimation (5 angles). In phase 1, the three angles represent the rotations of the hand palm (defined by 6 markers), around the x, y, and z axes, starting from the reference position, where the K plane (defined in Fig. \ref{fig:big_fig}) is aligned parallel to the W plane and the lines G and Q are parallel. As for phase 2, each finger in the model can perform flexion-extension and adduction-abduction at the metacarpo-phalangeal (MP) level, and flexion-extension at the interphalangeal (IP, ID) level. The four angles that best represent the spatial position of the three finger markers are derived using a sequential quadratic programming algorithm (MATLAB implementation; Maximal Function Evaluations=500; Finite Difference Type="forward"; function tolerance=10$^{-1}$; optimality=10$^{-6}$; step tolerances=10$^{-6}$; maximum iterations=1000; constraint tolerance=10$^{-6}$) \cite{nocedal1999numerical}. The algorithm is initialized with the anatomical rest angles of the hand \cite{bullock2012assessing}. All joint angles are equal to zero when the three finger markers lie on the K plane, on a line perpendicular to the Q line (Fig. \ref{fig:big_fig}). Positive angles indicate joint extensions and radial deviations. In phase 3, the thumb has the flexibility for flexion-extension and adduction-abduction at both the metacarpo-phalangeal and proximal interphalangeal (MP, IP) levels, with only flexion-extension at the distal interphalangeal (ID) level. The optimization algorithm, identical to the one in phase 2, computes the angles that best represent the position of the three thumb markers. All angles are equal to zero when those markers lie on the T plane, on a line perpendicular to the H line (Fig. \ref{fig:big_fig}). Joint extensions and radial deviations are indicated by positive angles. The angles for each finger and the thumb are calculated using separate optimization processes for enhanced computational efficiency. Phase 1 introduces no approximation errors, while phases 2 and 3 present a degree of discrepancy between estimated and actual positions. An average approximation error (distance between actual and estimated positions of the markers) of under 3 mm across all subjects was observed (mean: 2.35 mm, std: 1.69 mm computed over 1987466 frames). The IKA effectively performs a projection operation from a 3D space to a J-dimensional space, where J is the number of DoFs in the kinematic model (J=24). This model plays a crucial role as our neural network architecture, RPC-Net, was designed to take as input and return as output the joint angles of the hand rather than the 3D positions of the markers. This choice is motivated by the fact that muscle activation is proportional to joint angles rather than to finger endpoint position. A Forward Kinematic Algorithm (FKA) to translate angles to marker position, based on the same kinematic model, was developed as well.

\subsection{Data post processing}

The electromyographic and position data were post-processed using MATLAB (Release 2022b, The MathWorks, Inc., Natick, Massachusetts, USA) and VICON Nexus (VICON Nexus v2.11, Oxford Metrics plc, Oxford, UK). Additional code was written in Python using libraries that rely on BSD licenses. The objective of the post-processing procedure is to transform the data acquired into suitable inputs for RPC-Net. The post-processing procedure of the electromyographic signal consists of: 1) conversion of acquired signal from bit to volt, 2) removal of offset from signal, 3) rectification (computation of absolute value), 4) normalization (division by 5 mV) so that the signal is included between 0 and 1, 5) computation of the RMS of each channel over $w_l$=200 samples (97.7 ms) with a sliding window of $w_s$=25 samples. Given an electromyographic signal that is $L$ seconds long and $N$-dimensional, the output of the post-processing procedure is $l=floor(\frac{L*f_s-w_l}{w_s})+1$ samples long and $N$-dimensional. This procedure effectively sub-samples the signal from $f_s$=2.048 kHz to $\frac{f_s}{w_s}$=81.92 Hz through the computation of RMS. The post-processing procedure for the marker position data consists of: 1) moving average filtering (order 20), 2) projection of the positions of markers in 3D to a 24-dimensional joint-angle space through the IKA, 3) subtraction of rest angles, 4) normalization (addition of 150° and division by 240°) so that the signal is included between 0 and 1, 5) linear interpolation to match the sampling rate of the post-processed electromyographic signal ($\frac{f_s}{w_s}$). Given a $L$-second long signal, the output of the post-processing procedure is $l=floor(\frac{L*f_s-w_l}{w_s})+1$ samples long and J-dimensional.

\subsection{RPC-Net}

We devised RPC-Net, a neural network intended to convert multi-modal input into sample-wise predictions of the 24 hand joint angles. The architecture includes 24 individual sub-networks, each assigned to one joint. All sub-networks share the same input for each iteration and generate outputs that collectively represent the complete kinematic state of the hand. Each sub-network consists of a two-branched neural network, where one branch processes the electromyogram (EMG branch), and the other analyzes past joint angles (angle branch). These two branches converge into a root that returns an estimate for a single joint angle. The outputs from the 24 sub-networks are combined to create a full set of 24 joint angles, which also serve as input for the subsequent estimates, creating a recursive loop. Both input signals consist of multiple time points. The EMG input consists of a 0.78 s segment of the electromyographic signal preceding the instant at which the joint angle estimate is calculated. Of the 64 samples in this segment (given a sampling frequency of 81.92 Hz), only every fourth sample is processed for efficiency. This action effectively reduces the sampling frequency to 20.48 Hz. Consequently, the EMG branch input size becomes $I_E$=16 (samples) $\times$ 96 (channels) or 1536 inputs. The joint angle input incorporates the kinematic state for the 0.78 s interval leading up to the time point of the predicted position. Every eighth sample is chosen (diminishing the sampling rate to 10.24 Hz), and the signal encompasses J=24 channels. This results in an input size of $I_A$=8 (samples) $\times$ 24 (joints) or 192 inputs for the angle branch input. The output of the network is the sample-wise approximation of the 24 joint angles of the hand, determined at a frequency of 81.92 Hz. For myoelectrical control systems, delay is defined as the time difference between motion intention and output. While the input signal covers a 0.78 s window, this doesn't introduce any delay in the prediction process in addition to the inference time of the network. This is because RPC-Net performs the estimate of the sample immediately following the most recent EMG reading, corresponding to the onset of motor intention; earlier EMG values within the 0.78 s window improve prediction by providing previous context. The initial 0.78 s of a session may not be used for training or testing of the network due to the absence of sufficient earlier data. During the testing phase, joint angles produced by the RPC-Net undergo processing via a fourth-order low-pass Butterworth filter (with a cutoff frequency $f_c$=1 Hz) to eliminate high-frequency fluctuations and are subsequently mapped back into 3D space using the FKA. The structure of each branch is detailed in Fig. \ref{fig:NN_2}. The network was trained with the Adam optimizer (in its PyTorch implementation), learning rate=10$^{-5}$; $\varepsilon$=10$^{-3}$; $\beta_1$=0.9; $\beta_2$=0.99; batch size=10; loss criterion=MSELoss. The model was trained for 3 epochs. During the training of RPC-Net, the joint angle input data were from the recording, and the recursion loop was not used.

\subsection{Performance Indicators}
The performance of RPC-Net was assessed as its ability to estimate the position of the hand from the electromyogram. The performance was measured, independently for each subject, with to two indicators: Mean Pearson Correlation Coefficient (MPCC) and Mean Distance (MD), computed for the test trial only. MPCC is the mean of the individual Pearson Correlation Coefficients (PCC) obtained from comparing the actual and predicted joint angle value for each of the 24 DoFs considered, over the whole test trial. The median, first and second tertile of the PCCs were also computed along with the mean. MD is the mean, over the whole test trial, of the Weighted Fingertip Distance (WFD). For each time point estimated, WFD is defined as the average of: the distance between the position of the tip of the index finger (as recorded through VICON) and its estimate (computed by RPC-Net), the distance between the position of the tip of the middle finger and its estimate, and the distance between the position of the tip of the thumb and its estimate. The median, first tertile and second tertile of the WFD over time were computed. Computational efficiency was evaluated using Inference Time (IT). The operations in RPC-Net do not benefit from using a Graphics Processing Unit (GPU) during the prediction phase (forward passes), and the PyTorch library does not improve performance in this context. Thus, although GPU usage can accelerate the training phase, we found that performing matrix operations with the numpy library, rather than PyTorch, on a CPU yielded the best performance during testing. We calculated Inference Time as the average time, over 10$^5$ iterations, required by the numpy implementation of the network to perform a forward pass. Computation was performed on an Intel(R) Xeon(R) Platinum 8268 CPU (2.90 GHz).

\subsection{Comparison with other solutions}\label{sec:thrsltns}
We tested other state-of-the-art solutions for hand position estimation using our data set and compared their performance with that of RPC-Net. For each solution, we implemented a modified version to have a similar inference time to RPC-Net (ensuring a more fair comparison), to fit our input (96-channel HD-sEMG, 16 samples) and to have the same output as RPC-Net (24 joint angles). The solutions (identified with a code) considered are: 1) the Convolutional Neural Network proposed in \cite{olsson2019extraction_2}. The original implementation has 4 convolutional layers with 128, 64, 64, 64 output channels and 3 fully connected layers with 512, 512, 128 and 16 units respectively. Our modified version (CNN-Ols) has 4 convolutional layers with 128, 192, 192, 192 output channels and 3 fully connected layers with 1536, 1536, 384 and 24 units respectively. 2) The recurrent neural network proposed in \cite{Quivira2018}, originally with hidden size of 50 and 10 outputs, modified to have hidden size of 224 and 24 outputs (RNN-Qui). 3) The transformer architecture introduced in \cite{putro2024estimating}. The original implementation has 1024 as dimension of model, 4 transformer layers and 22 outputs. Our modified implementation (TRF-Put) has 96 as dimension of model, 12 transformer layers and 24 outputs. In our implementation, we omitted the ReLU function after the last fully connected layer in the regression block. The EMG input for RNN-Qui and TRF-Put was reshaped to a 16 (samples) $\times$ 96 (EMG channels) 2D array. For CNN-Ols, it was reshaped to a 16 (samples) $\times$ 6 (EMG electrodes rows) $\times$ 16 (EMG electrodes columns) 3D array. To compare the performance of RPC-Net with these alternative solutions, we performed a one-sided paired t-test between the performance of RPC-Net and the performance of TRF-Put, RNN-Qui, and CNN-Ols, across all subjects, for both MD and MPCC. The choice of the paired t-test was justified by the fact that it exhibits higher statistical power, under the assumption of normally distributed data, if compared to nonparametric tests. This test is widely regarded as a robust method for assessing differences in two paired populations \cite{ROSS2017433}. We opted for the one-sided version because we are assessing the superiority of RPC-Net in relation to other models. The condition of normality was verified with a Shapiro-Wilk test.

\subsection{Variations of RPC-Net} \label{sec:vrpcnet}
Additional simulations were run to assess the effect of computational-cost-related parameters on the performance of the network by means of different statistical tools. The effect of the inclusion of information about the previous joint state was assessed with one-sided paired t-tests (see \ref{sec:pr_j}), and so was the effect of using a single or multiple network for each joint (\ref{sec:s_o_24}). The choice of this statistical tool is justified in \ref{sec:thrsltns}. The condition of normality was verified with a Shapiro-Wilk test. For selected experimental conditions (\ref{sec:l_o_i}, \ref{sec:d_o_i} \ref{sec:s_o_24}), we compared the results observed with RPC-Net and RPC-Net-B (defined in \ref{sec:pr_j}). Since, in this case, data distribution normality could not be guaranteed, we relied on the one-sided Wilcoxon signed-rank test, a nonparametric alternative to the paired t-test \cite{ROSS2017433}. We opted for the one-sided version because we are assessing the superior performance of RPC-Net in relation to RPC-Net-B.

\subsubsection {Inclusion or exclusion of previous joint state as input to RPC-Net} \label{sec:pr_j}
To evaluate if information about previous joint state could enhance performance, we compared the performance the original RPC-Net architecture with that of a similar architecture that does not make use of information about previous kinematic state: RPC-Net-B (i.e., RPC-Net without the angle branch). The comparison was performed by means of a one-sided paired t-test between the results observed for of RPC-Net and RPC-Net-B across all subjects, for both MD and MPCC. 

\subsubsection {Length of input signal} \label{sec:l_o_i}
The length of the two signals given as input to RPC-Net was varied, keeping the sampling frequency unvaried, and thus modifying the number of samples in input. A different number of sampling points in input implied a change in network architecture, so that layer width was still a function of the number of input channels I{\tiny E} and I{\tiny A}, as shown in Fig. \ref{fig:NN_2}. The EMG and angle signal lengths were (rounded to the closest tenth of second) 0.8 s (original length), 0.7 s, 0.6 s, 0.5 s, 0.4 s, 0.3 s, 0.2 s, and 0.1 s. Linear regression analysis, effective in quantifying the strength and direction of a linear relationship, was used to analyze the effect on performance of the length of input signals. We tested the hypothesis that an increase in signal length correlates with an improvement in performance \cite{ROSS2017433}. The same assessment was performed on RPC-Net-B and the results compared through a Wilcoxon signed-rank test (96 coupled observations, 8 per subject, one each for RPC-Net and RPC-Net-B).

\subsubsection {Width of neural network} \label{sec:d_o_i}
The width of the neural layer of the EMG branch of RPC-Net was varied. Five widths were considered (in addition to the original one). The widths tested were: 512 (original width), 384 (E-1), 307 (E-2), 256 (E-3), 219 (E-4) and 192 (E-5) units, each identified by a code (E-$n$). The assessment was performed both on RPC-Net and RPC-Net-B and the results compared using a Wilcoxon signed-rank test (72 coupled observations, 6 per subject, one each for RPC-Net and RPC-Net-B).

\subsubsection {Number of electrodes used as input} \label{sec:n_o_e}
We assessed the performance of RPC-Net using a subset of EMG channels as input. Subsets were defined either on the proximo-distal axis or on the circumferential axis. This implied a change in network architecture so that the width of the layers was still a function of the number of input channels, as shown in Fig. \ref{fig:NN_2}. We tested the 12 combinations of row subsets and the 6 combinations of column subsets reported in Table \ref{table:ele_num} in addition to the original solution, which includes all rows and all columns. Column and row numbers are illustrated in Fig. \ref{fig:big_fig}.

\begin{table}[ht]
\caption{Input electrodes subsets}
\setlength{\tabcolsep}{3pt}
\renewcommand{\arraystretch}{1}

\begin{tabular}{|p{35pt}|p{35pt}||p{35pt}|p{35pt}||p{35pt}|p{35pt}|}
\hline
Code & Rows & Code & Rows & Code & Columns\\
\hline
A1 & 1-4 & C1 & 1 & D1 & Even\\
A2 & 1-2, 5-6 & C2 & 2 & D2 & Odd \\
A3 & 3-6 & C3 & 3 & F1 & 1-5-9-13 \\
B1 & 1-2 & C4 & 4 & F2 & 2-6-10-14\\
B2 & 3-4 & C5 & 5 & F3 & 3-7-11-15\\
B3 & 5-6 & C6 & 6 & F4 & 4-8-12-16\\

\hline

\end{tabular}
\label{table:ele_num}
\end{table}

\subsubsection {Single network for multiple joints or separate networks for each joint} \label{sec:s_o_24}
We developed two variations of RPC-Net (RPC-Net-W and RPC-Net-I) which, instead of 24 independent sub-networks (each corresponding to a joint angle), are made up of a single unit that takes the same inputs and returns 24 joint angles. In RPC-Net-I, the structure of the network is identical to that of a single branch in RPC-Net (see Fig. \ref{fig:NN_2}). RPC-Net-W is similar to RPC-Net-I, but the width of the layers was multiplied by a constant factor of 5 to match the inference time of RPC-Net, thus ensuring a more appropriate comparison in performance. The performance of these alternative architectures was assessed and compared with that of the original RPC-Net, by means of a one-sided paired t-test across all subjects, for both MD and MPCC. The assessment was performed both on RPC-Net and RPC-Net-B, for which analogous RPC-Net-I-B and RPC-Net-W-B were developed (i.e., RPC-Net-I-B is RPC-Net-I without joint angle input and RPC-Net-W-B is RPC-Net-W without joint angle input) and the results compared using a Wilcoxon signed-rank test (36 coupled observations, 3 per subject) with one observation each for RPC-Net (RPC-Net-W and RPC-Net-I) and RPC-Net-B (RPC-Net-W-B and RPC-Net-I-B).

\section{Results}

\subsection{Performance of RPC-Net} \label{sec:performance}

Fig. \ref{fig:big_pic} shows the trend of WFD values time, highlighting the discrepancy between RPC-Net estimates and actual data. We also show a similar measure, the Unweighted Marker Distance (UMD), computed with all the markers on the hand rather than with the index finger, middle finger and thumb alone. Table \ref{table:results} reports the MPCC and MD for all subjects. Fig. \ref{fig:results_small} shows an overlap of the angles as estimated by RPC-Net and their actual values, with joint codes defined in Fig. \ref{fig:kin_mod}. The results shown correspond to a one-minute interval from the test trial acquired for Subject S0. These results indicate that the estimate is consistently good across all joints. The best results in terms of PCC are observed for the flexion-extension of the wrist joint and the metacarpal-phalangeal joint of the four fingers. The inference time of RPC-Net is 6.20 ms, with a standard deviation of 0.01 ms.

\begin{table}[ht]
\caption{Performance indicators for RPC-Net across the twelve subjects}
\setlength{\tabcolsep}{3pt}
\renewcommand{\arraystretch}{1}

\begin{tabular}{|p{26pt}|p{120pt}|p{87pt}|}
\hline
Subject& 
MD (T1, T2, Med) in mm & MPCC (T1, T2, Med) in \% \\
\hline
S0 & 24.1 (18.9, 21.9, 25.3) & 80.4 (77.8, 82.2, 87.0) \\
S1 & 23.0 (17.5, 21.5, 26.0) & 81.8	(79.8, 86.8, 90.1) \\
S2 & 32.3 (25.2, 30.3, 35.3) & 76.3 (69.0, 83.2, 87.3) \\
S3 & 33.5 (23.6, 28.6, 36.7) & 77.6 (71.2, 83.3, 87.2) \\
S4 & 32.4 (22.6, 28.0, 33.6) & 75.7 (73.6, 77.1, 82.0) \\
S5 & 28.5 (20.4, 25.4, 31.8) & 83.0 (78.7, 85.7, 89.7) \\
S6 & 30.2 (22.2, 26.3, 32.7) & 78.8 (74.6, 81.6, 85.3) \\
S7 & 26.2 (19.6, 23.6, 29.1) & 79.5 (75.7, 83.0, 86.0) \\
S8 & 26.7 (19.3, 23.9, 29.4) & 66.0 (66.5, 71.6, 77.8) \\
S9 & 27.3 (17.5, 21.2, 26.2) & 76.9 (74.7, 79.3, 85.4) \\
S10 & 42.0 (26.1, 33.5, 43.7) & 77.8 (79.6, 83.9, 86.7) \\
S11 & 29.9 (21.8, 27.5, 33.7) & 79.3 (81.3, 84.5, 87.0) \\

\hline

\multicolumn{3}{p{251pt}}{Performance of RPC-Net for the twelve subjects considered in the study. The mean, first tertile (T1), second tertile (T2) and median (Med) are shown for each subject and for both performance indicators.}

\end{tabular}
\label{table:results}
\end{table}

\subsection{Variations of RPC-Net and comparison with other solutions}

Fig. \ref{fig:results_previous_state} to Fig. \ref{fig:separate_or_not} show the performance of RPC-Net if compared with other DL solutions and its own variants, measured by the previously defined indicators: MD, MPCC, and IT. Markers indicate actual indicators (i.e., the mean), intervals highlight the first and second tertiles, and a dot symbolizes the median. Statistics are included. For the one-sided paired t-test, we included the $t$ statistic and the corresponding p-value in the caption. For regression analyses, the caption included $\hat{\beta_{1}}$, $SE$, $t$, $p$, $R^{2}$ and adjusted $R^{2}$. In Fig. \ref{fig:results_depth_all}, Fig. \ref{fig:results_fcndepth_all}, and Fig. \ref{fig:separate_or_not}, the top and middle plot rows indicate architectures with or without information about previous position, respectively. The bottom row of plots compares the two conditions, using y-values from the top and middle rows to determine the position of a marker. Values on the y-axis in the top plot row are the x-axis in the bottom plot row, and values on the y-axis in the middle plot row are the y-axis in the bottom plot row. For the Wilcoxon signed-rank test we report the statistic $W$ (sum of the ranks of positive differences), and the p-value in the caption. Fig. \ref{fig:results_previous_state} compares the performance of RPC-Net, RPC-Net-B, TRF-Put, RNN-Qui, and CNN-Ols, referring to the experiments in sections \ref{sec:thrsltns} and \ref{sec:pr_j}. Statistical tests confirm the superior performance of RPC-Net against the other solutions. Another analysis (second row of plots), studies the impact of angle branch input signal length, revealing no significant influence on performance. Fig. \ref{fig:results_depth_all} depicts performance trends based on EMG branch input signal length for RPC-Net and RPC-Net-B, as detailed in section \ref{sec:l_o_i}. Longer signals predictably enhance performance. Fig. \ref{fig:results_fcndepth_all} examines the influence of the EMG branch layer width on performance for RPC-Net and RPC-Net-B, as defined in \ref{sec:d_o_i}. Visually, wider layers improve performance, though computational cost escalates quicker. Fig. \ref{fig:nr_electrodes} evaluates RPC-Net variants using a subset of electrodes, defined in section \ref{sec:n_o_e}. Fewer electrodes predictably lower performance. However, computational cost decreases significantly with fewer electrodes. Lastly, Fig. \ref{fig:separate_or_not} contrasts the performances of RPC-Net, RPC-Net-W, RPC-Net-I, and their RPC-Net-B counterparts, as described in \ref{sec:s_o_24}. RPC-Net outperforms RPC-Net-I (with a considerable difference in computational cost), but not RPC-Net-W. The results, as seen in the bottom plot row of Fig. \ref{fig:results_depth_all}, Fig. \ref{fig:results_fcndepth_all}, and Fig. \ref{fig:separate_or_not}, indicates that prior kinematic state information consistently improves performance.

\section{Discussion}

The most important consideration that can be inferred from the results is that RPC-Net is capable of translating electromyographic activity to hand position with good accuracy. This finding is of vital importance, as it demonstrates that a solution addressing user needs and allowing for natural-feeling control, while being computationally efficient, can achieve satisfying levels of performance; this opens new opportunities for the implementation of such solutions in a clinical setting. 

While other ML approaches yield results comparable to ours in EMG-based prosthetic control, these either control a limited number of DoFs \cite{Ameri_2019}\cite{nowak2022simultaneous}, rely on electrodes positioned too distally \cite{simpetru2022accurate}, or employ algorithms that demand substantial computational resources \cite{Bao2021A}\cite{Hu2015}. As for the second point, the use of electrodes positioned too distally (on the wrist or thumb) poses a challenge in applications, as many amputees lack these parts of the limbs \cite{cordella2016literature}. The results indicate that all the necessary information to determine the current kinematic state of the hand can be derived from the electromyographic activity in the proximal portion of the forearm, even for joints with distal or intrinsic actuators. Although the adequacy of forearm electromyogram for prosthetic control has been demonstrated \cite{mendez2021}, no study yet demonstrated the feasibility of all-DoF control using electromyographic activity acquired solely from the proximal forearm. On the computational front, algorithms that demand extensive resources necessitate advanced hardware like GPUs, which are impractical to embed into a prosthetic device, thereby limiting their applicability in clinical settings. In this regard, the results point towards another novel finding: high-quality regression can be achieved using simple fully connected neural architectures, rather than complex convolutional structures, a result demonstrated by the performance of RPC-Net if compared to our implementation of the solutions proposed by Quivira et al. \cite{Quivira2018}, Olsson et al. \cite{olsson2019extraction_2}, and Putro et al \cite{putro2024estimating}. A simple architecture implies a reduced computational cost and makes RPC-Net potentially embeddable in a microprocessor of a prosthetic device.

Incorporating information about the previous kinematic state significantly contributes to the performance of our solution. The results show that the improvement in performance given by the inclusion of information about previous state is consistent across subjects and across conditions. In real-life situations, this feature could offer a stabilizing mechanism to counteract the stochastic nature of electromyographic signals. The results illustrate the resilience of RPC-Net against reductions in EMG electrode numbers, shorter input signals, and other conditions that could lower computational cost. The findings indicate that while additional input data typically enhances estimation quality, the improvement doesn't correspond proportionally to computational cost. For instance, reducing the number of electrodes from 96 to 32 only increased fingertip distance by 5 mm and reduced MPCC by 10\%, yet resulted in a 75\% reduction in computation time. Hence, a control solution based on RPC-Net could offer substantial flexibility and could be implemented with fewer electrodes or a shorter input signal, implying that the computational cost of the solution could be further minimized with minimal performance loss, making it highly adaptable to clinical settings. Previous studies have explored the adequacy of fewer electrodes for control \cite{Amma2015} \cite{ROJASMARTINEZ201333}, but none for these many DoFs. 

The experiments described in this work were conducted entirely offline and on healthy subjects. For RPC-Net to be properly validated for controlling upper limb prostheses, its performance must be demonstrated in conditions more closely resembling the actual target case. To achieve this, the next steps in our research include implementing RPC-Net in real-time, integrating it for the control of an existing, state-of-the-art prosthetic device, and applying it to amputees. Finally, once the solution is successfully applied to amputees, a subsequent study would be necessary to assess their satisfaction and to determine the potential positive impact of a prosthetic device controlled by RPC-Net.

\section{Conclusion}

In this study, we introduced and extensively evaluated RPC-Net, a method for translating electromyographic activity into hand position using a computationally efficient neural network architecture. The results of the study show that RPC-Net can perform better than state-of-the-art methods with a similar computational cost and without requiring advanced computer resources. This signifies a crucial development, making the system more adaptable and potentially more accessible in clinical settings. A key feature that substantially contributes to the performance of the system was incorporating information about the previous kinematic state. Furthermore, our solution demonstrated robustness against changes in input signal length and the number of EMG channels used, reinforcing its flexibility and adaptability. The development and success of RPC-Net have significant implications for biomedical research and practical applications. It offers a new avenue for prosthetic control that feels more natural, is computationally efficient, and can be flexibly adapted to different clinical settings and patient needs. This progress has the potential to improve the quality of life for individuals with limb loss, pushing forward biomedical research and its applications. Our future research will focus on taking the necessary steps to implement RPC-Net as a control solution for prosthetic devices.

\section{Acknowledgment}

The authors would like to thank the members of ONIG in Oxford and of LISiN in Turin for the valuable guidance provided during the writing of this article and to acknowledge the use of the University of Oxford Advanced Research Computing (ARC) facility in carrying out this work \cite{arc_zenodo}. Fig. \ref{fig:big_fig} and Fig. \ref{fig:kin_mod} were created with BioRender.com.


\end{document}